\documentclass[prl,twocolumn,aps,showpacs,amsmath,amssymb]{revtex4-1}

\usepackage{graphicx}
\usepackage{epstopdf}
\usepackage{dcolumn}
\usepackage{bm}
\usepackage{amsmath}
\usepackage{epsfig}
\usepackage{float}
\usepackage{epstopdf}

\begin{document}
	
	\title{Discovery of Terahertz Second Harmonic Generation from Lightwave Acceleration of Symmetry--Breaking Nonlinear Supercurrents}
	
	
	\author
	{C. Vaswani$^{1\dagger}$, C. Sundahl$^{2\dagger}$, M.~Mootz$^{3\dagger}$, D. H. Mudiyanselage$^{1}$, J.~H.~Kang$^{2}$, X. Yang$^{1}$, D. Cheng$^{1}$, 
		C. Huang$^{1}$, R.~H.~J.~Kim$^{1}$, Z. Liu $^{1}$, L. Luo$^{1}$, 
		I.~E.~Perakis$^{3}$, C.~B.~Eom$^{2}$ and J. Wang$^{1}$}
		\affiliation{$^1$Department of Physics and Astronomy and Ames Laboratory-U.S. DOE, Iowa State University, Ames, Iowa 50011, USA. 
		\\$^2$Department of Materials Science and Engineering, University of Wisconsin-Madison, Madison, WI 53706, USA
		\\$^3$Department of Physics, University of Alabama at Birmingham, Birmingham, AL 35294-1170, USA. }
	
	
	\date{\today}

	\begin{abstract}
	We report terahertz (THz) second harmonic generation (SHG) in superconductors (SC) with inversion symmetric equilibrium states that forbid even-order nonlinearities. Such SHG signal is observed in single-pulse emission by periodic driving with a multi-cycle THz electric field tuned below the SC energy gap and vanishes above the SC critical temperature.  We explain the microscopic physics by a dynamical symmetry breaking principle at sub-THz-cycle by
using quantum kinetic modeling of the interplay between strong THz-lightwave nonlinearity and pulse propagation. The 
 resulting non-zero integrated pulse area inside the SC drives lightwave nonlinear supercurrents due to sub--cycle Cooper pair acceleration, in contrast to
 d.c.-biased superconductors,   
which can be controlled by the bandstructure and the THz pump field.   

\end{abstract}

\maketitle
The determination and understanding of symmetry breaking in superconducting states has been a central theme in condensed matter physics that remains challenging. A recent example is SHG at optical frequencies that is actively explored in cuprates and other inversion-symmetry-breaking superconductors \cite{XU19,ZHA17}. Such studies reveal that, in addition to the underlying crystal structure, the quantum order itself can also lead to non-trivial SHG signals.  In contrast to high energy optical excitation, the advent of intense few--cycle THz pulses has opened new  opportunities for exploring fundamental nonlinear physics and broken symmetry states \cite{KAM13}. 
Multi-cycle phase-locked THz pulses tuned below the pair-breaking energy gap 2$\Delta_\mathrm{SC}$ minimally perturb SC states. In contrast, optical pumping tends to destroy SC order by heating the quasi-particles (QPs) \cite{YAN19A}. In addition, while the carrier-envelope phase-$unlocked$ pulses used for optical pumping are sensitive to SHG, they are not suitable for identifying sub-cycle lightwave modulation effects that relate to the oscillating pump $E$-field. THz-induced nonlinear effects in superconductors have been of interest lately, e.g., third harmonic generation (THG) revealing Higgs \cite{MAT14, CEA16, Manske, CEA19, Kemper, chu19}, Leggett modes \cite{GIO19} and strip phases \cite{Rajasekaran}, a single--cycle THz-driven gapless QP fluid with vanishing scattering \cite{YAN18}, and the observation of higher harmonics (HH) in coherent pump-probe responses \cite{YAN19B}. However, forbidden THz-induced SHG, or T-SHG, 
from single--pulse excitation of SC states has not been observed so far.  

SHG may be observed in SCs with an additional inversion symmetry breaking order parameter coming, e.g., from pseudo-gap, magnetic, charge, or  lattice coupled orders.  However, the {\em spontaneous} coherence between Cooper pairs ($\mathbf{k}\uparrow$, $-\mathbf{k}\downarrow$) in a simple BCS ground state does not support SHG, due to the inversion symmetry. Nevertheless, {\em driven} coherence by strong acceleration of macroscopic Cooper pair center-of-mass (CM) motion can transiently break the equilibrium inversion symmetry without pair breaking, via  a periodically modulated superfluid momentum, $\mathbf{p}_s(t) 
\propto \int_{-\infty}^{t}\mathrm{d}\tau\,\mathbf{E}_{eff}(\tau)$.
Such time--dependent preferred direction can be introduced 
 by phase-locked THz electric field pulses tuned below the 2$\Delta_\mathrm{SC}$ gap, which induce an 
effective local electric field $\mathbf{E}_{eff}(\tau)$  determined by the electromagnetic fields and by spatial gradients of the  chemical potential and scalar fields. 
Fig.~1(a) illustrates the quantum dynamics of the BCS state driven by an a.c. field, arising from precession of  Anderson pseudospins (PSs) mapped onto the Bloch sphere. In this Bloch picture, the PSs respond to a pseudomagnetic field controlled by THz driving, whose $x$/$y$-components (transverse) are given by the complex SC order parameter, whereas its $z$-component (longitudinal) is determined by the bandstructure. 
Cooper pair lightwave acceleration can
non--adiabatically drive a {\em supercurrent--carrying}
transient macroscopic state, with oscillating condensate momentum $\mathbf{p}_s(t)$ (red line, Fig.~1(a)),  consisting of pairs ($\mathbf{k}+\mathbf{p}_\mathrm{s}(t)\uparrow$, $-\mathbf{k}+\mathbf{p}_\mathrm{s}(t)\downarrow$)  \cite{YAN19B}. 
The resulting light-induced nonlinear supercurrent flow 
breaks the equilibrium symmetry that gives rise to PS dynamics. 
Such PS oscillations have manifested themselves in the forbidden third-harmonic peaks observed in the two--pulse pump--probe spectra 
of sufficiently clean Nb$_3$Sn SCs \cite{YAN19B}. However, T-SHG in single--pulse nonlinear emission, a hallmark for the driven broken-symmetry state, has not been observed, raising some questions about the interpretation of the pump-probe signals in Ref.\cite{YAN19B}. 
More importantly, this lightwave current tuning should be distinguished from the d.c. biased SCs from an application of electrodes \cite{current}. 
Two key questions must be addressed to fully underpin the microscopic physics: (i)  how can an {\em asymmetric} ac electric field pulse with non-zero pulse area $\int_{-\infty}^{\infty}\mathrm{d}\tau\,\mathbf{E}(\tau)\neq 0$, i.e., a zero-frequency (dc) component, 
be generated in a nonlinear condensate medium? (ii) Is the Nb$_3$Sn bandstructure, where flat bands close to the Fermi level result in a large DOS, responsible for enhancing the lightwave supercurrent?

\begin{figure}[!tbp]
	\centering
	\includegraphics[scale=0.4]{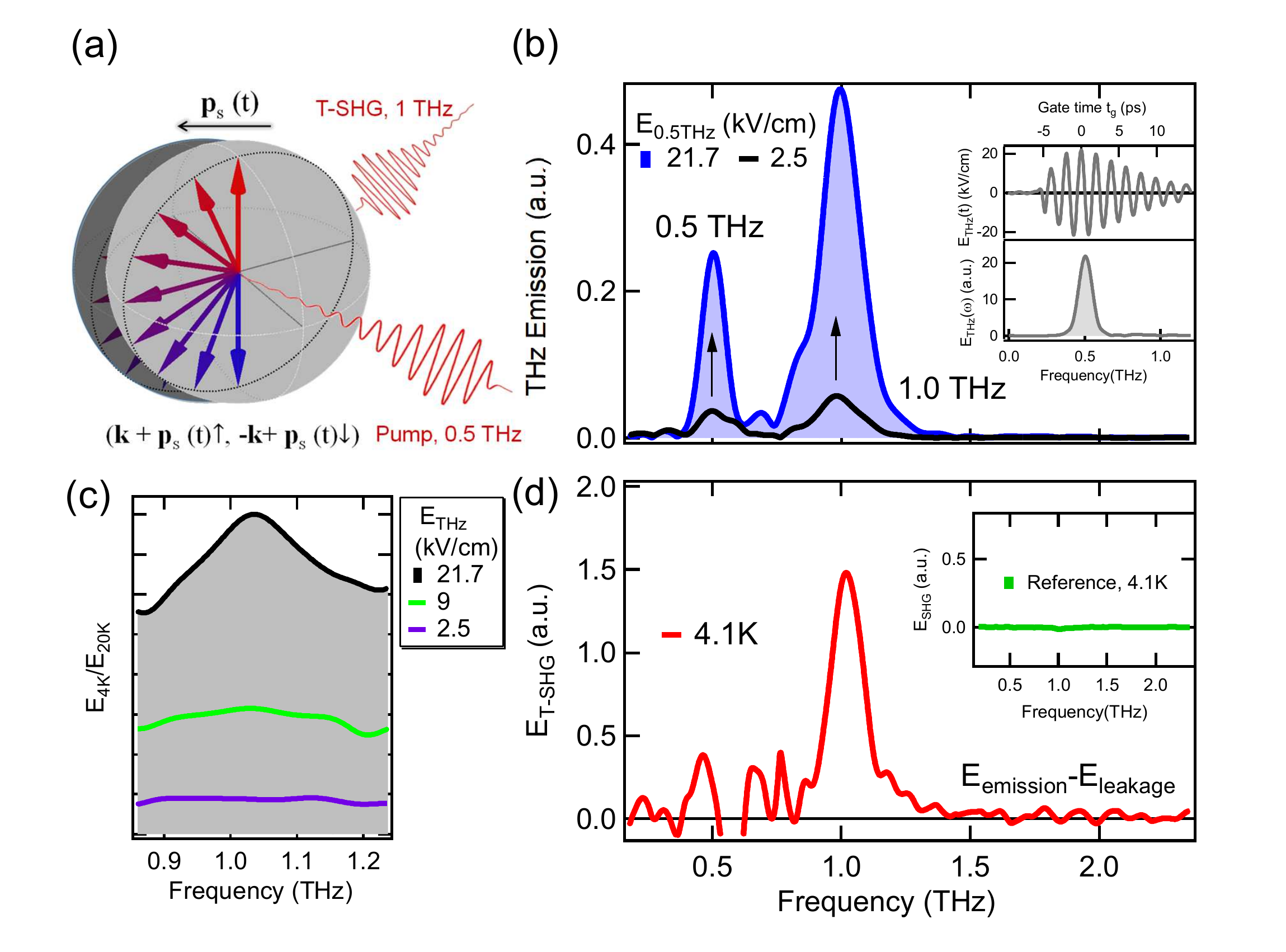}
	\caption{(a) Illustration of THz SHG generation by Anderson pseudospins driven by THz lightwave acceleration of superfluid momentum, $\mathbf{p}_s(t)$. 
		(d) THz emission for two THz E-field strengths, 21.7\,kV/cm(blue) and 2.5\,kV/cm(black). Inset: Representative 0.5THz multicycle phase-locked THz pulse and spectrum. (c) THz emission at 4.1K normalized to the emission at 20K for various THz E-field strengths (traces offset for clarity). (d) The THz SHG contribution to the measured THz emission signals after subtracting the pump leakage (main text). Inset: THz SHG emission is absent from the substrate after accounting for the pump leakage.}
	\label{Fig1} 
\end{figure}

In this letter, we demonstrate single--pulse nonlinear emission SHG below the SC critical temperature (T$_\mathrm{c}$). Such SHG is forbidden by the equilibrium symmetry of Nb$_3$Sn. We present nonlinear quantum kinetic calculations, based on gauge invariant density matrix equations of motion, which are in excellent agreement with the experimental results. We thus underpin the photogeneration of a broken symmetry nonlinear supercurrent arising from THz lightwave propagation for a nonlinear s--wave BCS medium. More importantly, our results describe a microscopic mechanism for photogenerating low--frequency components in the forward- and backward-traveling THz electric fields in the presence of PS nonlinearity, controlled by the bandstructure and THz field. 

Our sample consists of a 20~nm thick Nb$_3$Sn film grown by magnetron sputtering on an Al$_2$O$_3$ substrate. T$_\mathrm{c}\sim 16$~K and SC gap 2$\Delta_\mathrm{SC}\sim 4.5$~meV \cite{YAN19A, peter}.  Information on  sample growth procedures and equilibrium electrodynamics is discussed elsewhere~\cite{YAN18,YAN19B}. 2~W, 35~fs pulses from a Ti:Sapphire based regenerative amplifier were used to generate broadband quasi single-cycle THz pulses from a LiNbO$_3$ crystal, via a tilted-pulse-front scheme~\cite{YAN18}, with peak electric field $\sim$1000 kVcm$^{-1}$ and broad bandwidth $\sim$0-3 THz. Narrowband multi-cycle THz pulses at 2.1~meV (0.5~THz) were then obtained by using narrow band-pass filters. The peak electric field, E$_{\mathrm{0.5THz}}\sim$20kVcm$^{-1}$, is shown in the inset of Fig.~1(b) along with the pulse spectrum. 
For the results presented below, 
a 4.2~meV (1.0~THz) band pass filter was placed after the sample to block the fundamental beam and extract the nonlinear signal. Three wire-grid polarizers inserted in the THz beam path were used to control the $E$-field strength without changing the polarization of the THz driving field. The THz emission was detected by using electro-optic sampling in a 1\,mm ZnTe crystal.  

Figs.~1(b)-1(c) show the observation of T-SHG emission at 1.0~THz for various  field strengths E$_{\mathrm{0.5THz}}$.   
These THz emission signals are, however, a mixture of both linear (THz pump background) and nonlinear responses, since it is not possible to completely filter out the pump.  
This is evident in Fig.~1(b), which plots the THz emission from our sample for THz pump $E$-field strengths of $E_\mathrm{0.5THz}=$21.7~kV/cm and 2.5~kV/cm. A signal at 0.5~THz  is clearly visible even after the 1.0~THz filter is placed after the sample, due to residual leakage of 0.5~THz radiation through the 1.0~THz filter. Likewise, a  portion of the signal at 1.0~THz should arise from leakage of 1.0~THz radiation from the 0.5~THz filter placed in the pump’s path before the  excitation  to narrow the broadband THz pump spectrum.
To extract the nonlinear contribution to the T-SHG emission coming from the SC order, Fig. 1(c) shows the $\sim$1.0~THz signals at 4~K normalized to the normal state value measured at 20~K, i.e., $E_{4.1K}/E_{20K}$, for various  field strengths E$_{\mathrm{0.5THz}}$=21.4, 9, 2.5 kVcm$^{-1}$. For high $E$-fields, the emission  at 1.0~THz shows a   strong nonlinear field dependence, which diminishes at low $E$-fields. While the 2.5~kV/cm trace shows a temperature-independent SHG signal,  attributed to pump leakage, the observation of nonlinear behavior indicates a contribution at 1.0~THz from the SC state. For the 21.4 kVcm$^{-1}$ trace, this nonlinear SC contribution becomes bigger than the pump leakage contribution also observed in the normal state. We thus attribute the temperature-dependent, nonlinear SHG emission for 21.4 kV cm$^{-1}$ to a dynamically generated T-SHG effect  elaborated below. 

To demonstrate  the second order nature of the nonlinear THz emission in Fig.~1(c), we subtract the pump leakage contribution to the measured THz transmission with the  following procedure. The pump leakage contribution for a given $E$-field strength can be estimated  by scaling the low field data at 2.5~kV/cm according to the leakage ratio obtained from the pump polarizer angle. We thus obtain the results in Fig. 1(d) for 21.7~kV/cm at 4.1~K.         
This contribution to the T-SHG emission shows a well-defined resonance at 2$\omega_\mathrm{pump}$.  
Note that there is no measurable T-SHG signal from the sapphire substrate, as seen from Fig.~1(d) (inset).
Fig.~2(a) shows the $E$-field dependence of the  T-SHG signal extracted from the measured emission as above. The peak of this T-SHG contribution is plotted against the square of the normalized $E$-field strength $(E_\mathrm{THz}$/$E_\mathrm{max})^2$ in Fig.~2(b). The observed $E$--dependence  is well reproduced by a linear fit, as expected for a second--order non-linear optical process, i.e., our 1THz signal is proportional to $E^{2}_\mathrm{pump}$. Note that any residual contribution from filter leakage should be linear in $E$. This   corroborates our assignment of the  nonlinear T-SHG effect observed  below T$_c$ to the SC order. 

\begin{figure}[!tbp]
	\centering
	\includegraphics[scale=0.4]{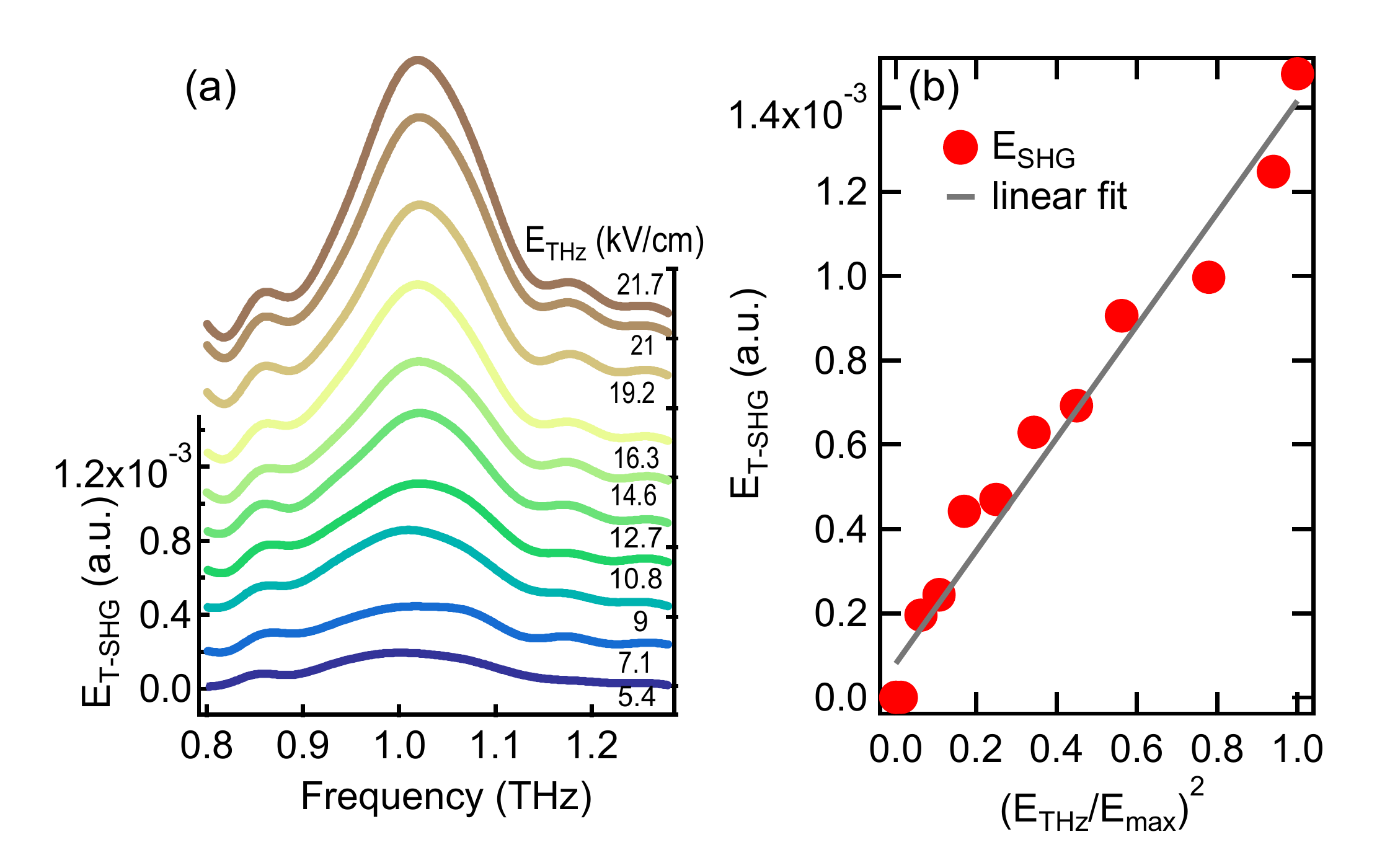}
	\caption{(a)THz SHG signals for various E-field strengths at 4K (traces offset for clarity). (b) THz SHG at 1\,THz as function of the square of the normalized E-field strength. The grey line shows a linear fit to the data.}
\end{figure}  
 
Figure~3(a) shows the strong temperature dependence of the above T-SHG emission. To  accurately extract this SHG  temperature dependence, we must account for the change in THz transmission due to temperature dependence of the $E$-field transmittance.  
This was done by normalizing the measured THz emission signals $E_{T}/E_{20K}$, Fig. 1(c), at each temperature $T$ by the transmittance (T$=E_\mathrm{Sample}/E_\mathrm{Reference}$) at that temperature. 
The resulting quantity, $(E_{T}/T_{T})/(E_{20K}/T_{20K})$, should  describe the temperature dependence of the T-SHG contribution. 
This T-SHG resonance at 2$\omega_\mathrm{pump}$ vanishes, with a fairly constant lineshape, at the critical temperature T$_c \sim$16\,~K, Fig. 3(b) (dashed line). 
The measured temperature dependence follows that of the SC order parameter, which indicates that the origin of the forbidden T-SHG behavior is 
light--induced  condensate motion, rather than some other contribution such as surface nonlinearity that contributes at all temperatures.  

\begin{figure}[!tbp]
	\centering
		\includegraphics[scale=0.4]{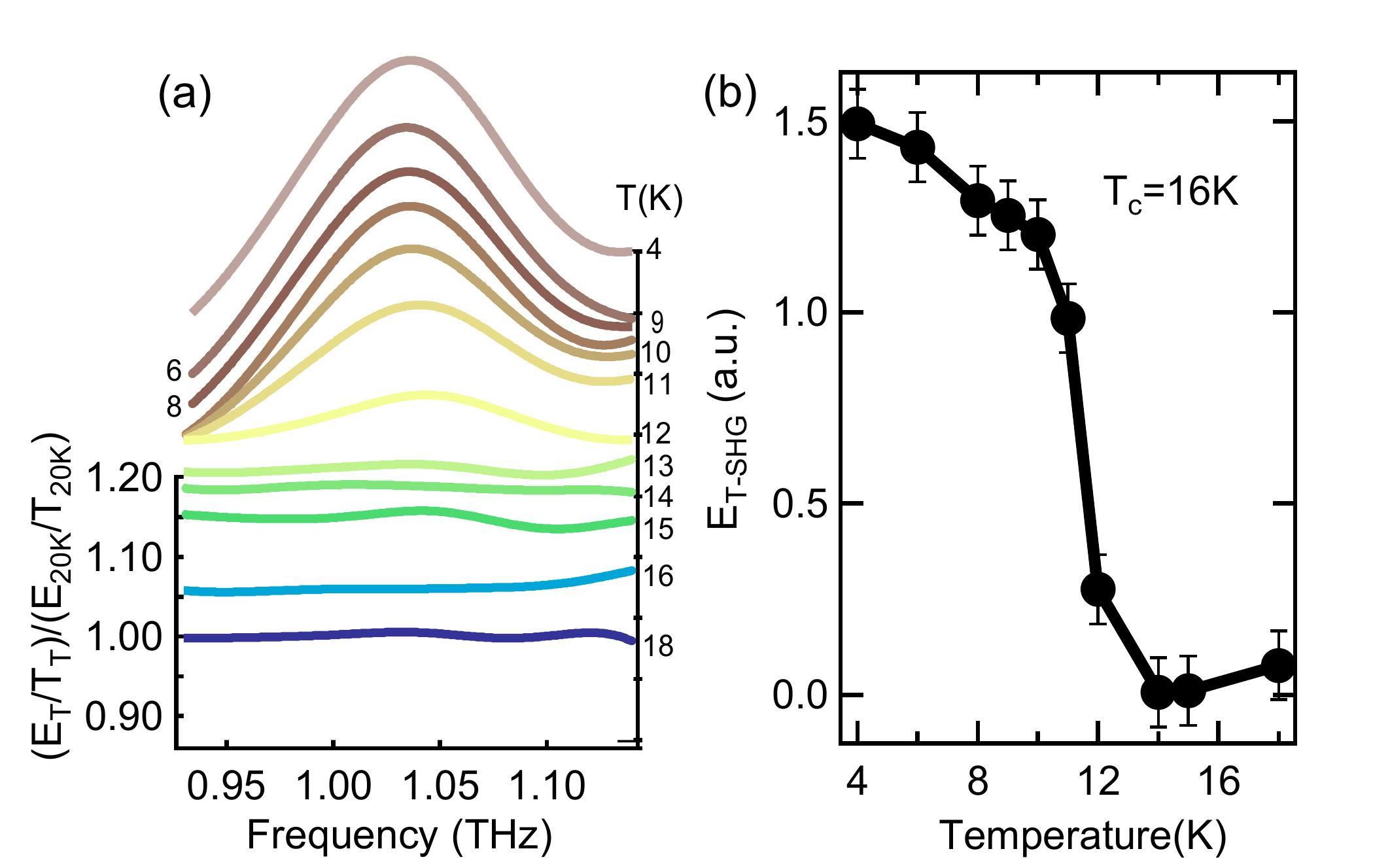}
		\caption{(a) THz-SHG signals scaled to the E field transmittance at various temperatures normalized by the 20K data (see text, traces offset for clarity). (b) Temperature dependence of the Integrated spectral weight of THz SHG signals.}
\end{figure}

To model our proposed mechanism for nonlinear lightwave supercurrent photogeneration, 
we extend previous studies of quantum transport~\cite{Stephen1965,Wu2017,YAN19B} and HH generation (HHG) ~\cite{Aoki2017,Cea2018} in SCs by
including the self-consistent interaction of the SC system with the propagating electromagnetic field (Supplemental Material). The sub--cylce  time--dependence is  described in a gauge-invariant way by generalizing the treatment of analogous ultrafast quantum kinetic transport effects in semiconductors to include the off-diagonal long range order \cite{Haug}. We thus derive gauge-invariant SC Bloch equations ~\cite{YAN19B} after subsequent gradient expansion of the spatial fluctuations \cite{Haug}. Together with Maxwell's equations, we thus describe the dynamical interplay of three different THz-light induced ultrafast effects: (1) Lightwave nonlinear acceleration of the Cooper-pair condensate, (2) Anderson pseudo-spin nonlinear precession, and (3) THz lightwave propagation inside the SC thin film geometry. The latter propagation effects are required for photogeneration of a dc supercurrent component
in the presence of SC nonlinear response. The latter is due to both THz-light induced condensate acceleration and pseudo-spin precession  and affects the interference between incident and reflected propagating waves. Here we do not consider additional effects resulting from the coupling of the SC order to the competing martensitic lattice order in the studied Nb$_3$Sn SC \cite{YAN18}.

\begin{figure}[!tbp]
	\centering
	\includegraphics[width=1.0\columnwidth]{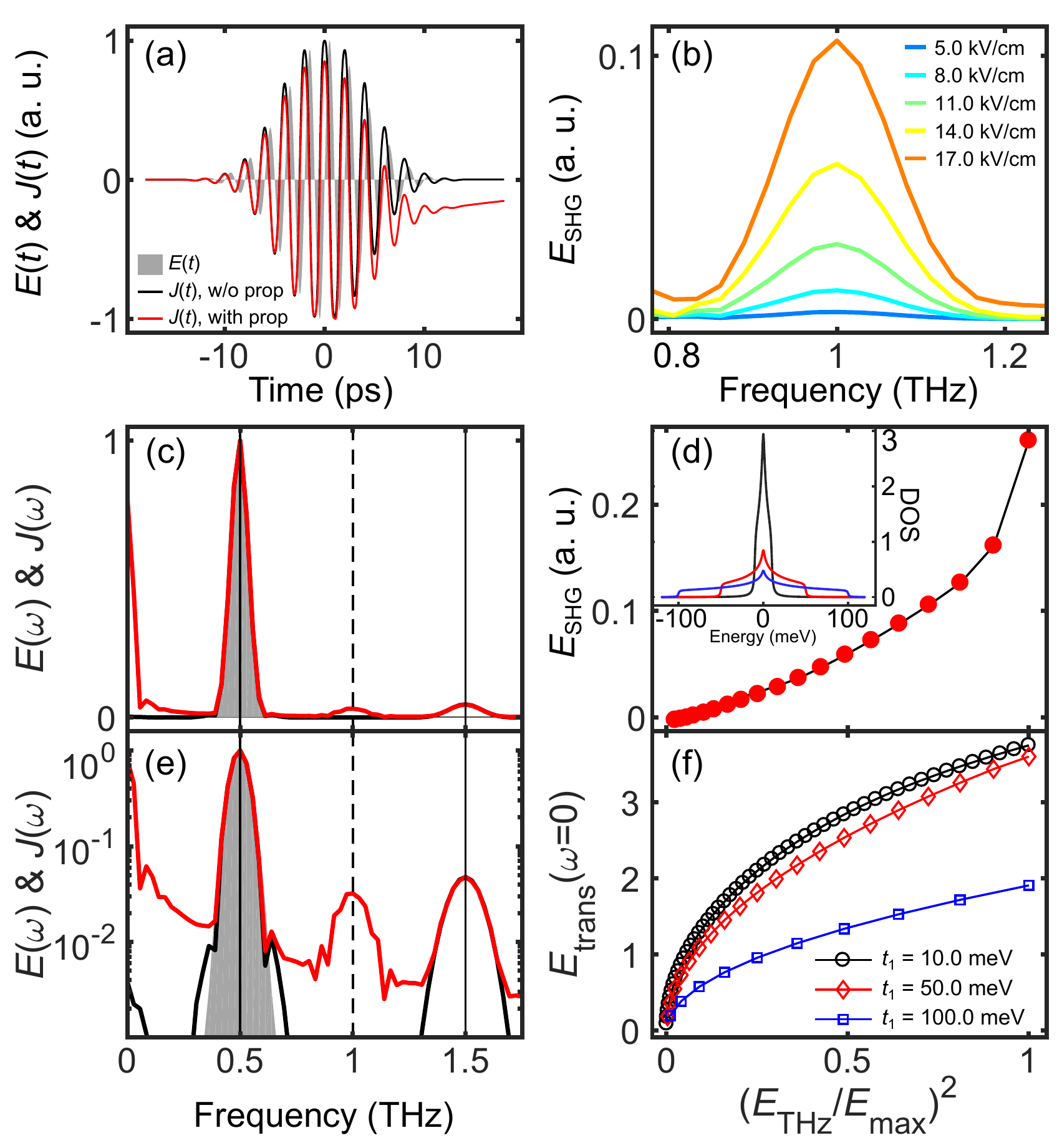}
	\caption{Gauge-invariant quantum kinetic simulation of dynamical symmetry breaking and nonlinear supercurrent photogeneration by THz lightwave propagation and interference effects. (a) Dynamics of THz-light induced nonlinear supercurrent $J(t)$, calculated without (black line) and with propagation effects (red line), together with the representative 0.5~THz pump oscillating electric field used in the calculations (shaded area). (b) Calculated THz SHG for various $E$-field strengths. Without the propagation effects, this signal is zero; (c), (e) Calculated nonlinear spectra over a range of frequencies, in linear and semi-logarithmic scale; the linear and third harmonic emission peaks are indicated by vertical solid lines, while SHG is denoted by vertical dashed line. (d) Calculated non-perturbative THz SHG at 1~THz as a function of the square of the normalized $E$-field strength. (f) Fluence dependence of the zero-frequency component of the transmitted nonlinear $E$-field for three different electron hopping strengths $t_1$ that characterize the flatness of the electronic bands. Inset: DOS for the different used $t_1$.}
\end{figure}

Based on Maxwell's equations, any physical source of electromagnetic waves cannot contain a zero-frequency dc component~\cite{Kozlov2011}: $\int_{-\infty}^{\infty}\mathrm{d}t\,E_\mathrm{THz}(t)= 0$. However, this does not apply to reflected and transmitted electric  field pulses after interaction with a nonlinear medium~\cite{Kozlov2011}. A dynamical broken-symmetry dc supercurrent is photogenerated  via the following steps.
First, THz excitation of the SC with $E_\mathrm{THz}(t)$ creates a nonlinear ac supercurrent $J(t)$, which then generates an electric field that interferes with the forward- and reflected backward-traveling THz electric fields inside the nonlinear SC.  Such interference results in  time-asymmetric reflected ($E_\mathrm{ref}(t)$) and transmitted ($E_\mathrm{trans}(t)$) electric field pulses with $\int_{-\infty}^{\infty}\mathrm{d}t\,E_{\mathrm{ref},\mathrm{trans}}(t)\neq 0$ inside the SC. The static component of the THz-light induced current is the source of a zero-frequency component of the sub-pulses~\cite{Kozlov2011}. The strength of this photogenerated component of the reflected and transmitted electric fields depends on the THz-light indued SC nonlinearities. The latter are controlled by the effective local field spectral and temporal properties, as well as by the intensity of the applied pump $E$-field and the bandstructure, discussed below. In the second step, the SC interaction with the above dynamically generated asymmetric effective electric field pulse  breaks the equilibrium inversion symmetry and induces a Cooper-pair condensate flow. The latter can persist well after the pulse assuming  weak photocurrent relaxation. 

Figure~4(a) illustrates the calculated supercurrent photogeneration via THz pulse propagation inside the SC system. The external pump electric field $E_\mathrm{THz}(t)$ (shaded area)  is shown together with the photo-induced current $J(t)$ resulting from our calculation without (black line) and with propagation effects (red line). We used $E_\mathrm{THz}(t)=\tilde{E}(t)\,\mathrm{sin}(\omega_\mathrm{pump} t)$ with Gaussian envelope $\tilde{E}(t)$, which satisfies $\int_{-\infty}^{\infty}\mathrm{d}t\,E_{\mathrm{THz}}(t)= 0$.   $\omega_\mathrm{pump}=2.1$~meV is well  below the SC gap $2\Delta_\mathrm{SC}=4.5$~meV. The pulse duration $\sim 20$~ps is similar to the experimental pump pulse (inset Fig.~1(c)). The photoinduced supercurrent resulting from our calculation including propagation effects (red line) remains finite after the pulse, in contrast to the result without THz lightwave propagation (black line).  This calculation demonstrates that a significant dc component of the photocurrent can be induced by THz lightwave propagation inside a SC thin film as discussed above. The calculated decay of this photoinduced dc supercurrent after the pulse here comes from radiative damping and results from self-consistent coupling between the current and  laser field. 

The predicted inversion-symmetry breaking in the non-equilibrium moving condensate is experimentally detectable via high harmonics emitted at equilibrium-symmetry-forbidden frequencies. This is demonstrated in Figs.~4(c) and (e), where the spectra of the pump electric field and the currents of Fig.~4(a) are shown in linear and semi-logarithmic scale, respectively. The spectrum of the current resulting from our calculation including THz lightwave propagation (red line) exhibits an equilibrium-symmetry forbidden SHG generation (vertical dashed line), as well as a pronounced zero-frequency component. These contributions are in addition to the equilibrium--symmetry--allowed linear and third harmonic emission (vertical solid lines) reported before. 
In comparison, the spectrum of the current resulting from our calculation without THz light-wave propagation effects (black line) shows only odd harmonics, as expected. We conclude that THz-light induced nonlinearities, together with THz-lightwave propagation inside the SC system, can induce a Cooper-pair condensate flow which manifests itself in equilibrium-symmetry-forbidden SHG. 

Figure~4(b) presents the calculated SHG spectra of the transmitted electric field for five different electric field strengths. A resonance emerges at the SHG frequency of $1.0$~THz with increasing pump fluence. This behavior is in agreement with the experimental observations in Fig.~2(a).  The fluence dependence of the SHG signal (Fig.~4(d)) shows a linear dependence as a function of the square of the normalized $E$-field $(E_\mathrm{THz}/E_\mathrm{THz,max})^2$ at low electric field strengths, in agreement with the experimental results (Fig.~2(b)). Deviations from this behavior at higher pump fluences emerge when the SC order parameter is significantly quenched by the THz $E$-field. Here, the interplay of dynamical symmetry breaking due to $\mathbf{p}_\mathrm{s}(t)$ and HHG nonlinearities enhanced by the pairing interaction produce strongly nonlinear quantum dynamics beyond perturbation expansions~\cite{YAN19B}.

To explore why nonlinear supercurrent phtogeneration by THz light-wave propagation is very effective in Nb$_3$Sn SCs, we study the effect of the bandstructure on the photogeneration of the zero-frequency component of the transmitted electric field. For this we use a square lattice nearest-neighbor tight-binding model, $\varepsilon(\mathbf{k})=-2\,t_1[\mathrm{cos}(k_x\,a)+\mathrm{cos}(k_y\,a)]+\mu$, with nearest-neighbor hopping strength $t_1 > 0$, lattice constant $a$, and band-offset $\mu$.  We characterize the effects of the bandstructure by the density-of-states (DOS) close to the Fermi surface. A small electronic hopping parameter $t_1$ corresponds to flatter band  dispersion and large DOS around the Fermi surface;  large $t_1$ yields a small DOS. 
Figure~4(f) shows the static component of the transmitted electric field $E_\mathrm{trans}(\omega=0)$ as a function of normalized electric field strength for three different DOS (inset Fig.~4(d)) obtained by changing $t_1$. The photoinduced supercurrent grows with increasing DOS at the Fermi surface, which shows that dynamical inversion symmetry breaking is most effective in SCs  with  small band dispersion (large DOS) close to the Fermi surface. This is the case in the Nb$_3$Sn superconductors studied here ~\cite{Sadigh1998,PADUANI2009}.



In summary, we describe a microscopic  mechanism of dynamical symmetry breaking that manifests itself experimentally via THz second harmonic generation (SHG) forbidden by the equilibrium pairing symmetry and absent above the SC transition temperature. Our experimental observation of this predicted SHG signal is a hallmark for symmetry breaking condensate flow induced by strong THz field-SC coupling in a thin film  geometry. Our theory-experiment results reinforce a universal quantum control concept of how oscillating THz electromagnetic field pulses can be used as an alternating-current bias to photogenerate sub-cycle dynamical spatial symmetry breaking in quantum materials. This scheme can be extended to other systems to such as topological matter \cite{more1}, magnetism \cite{more2} and unconventional superconductivity \cite{more3, more4, Patz2018} by dynamically tuning the symmetry. For example, periodically driven symmetry breaking current photogeneration can apply to high-Tc cuprates for probing spatial periodic modulation of Cooper pairing, as in a pair density wave state.


\begin{thebibliography}{99}
	
	\bibitem{XU19} T. Xu, T. Morimoto, and J. E. Moore, 
	arXiv: 1908.10476 (2019).
	
	\bibitem{ZHA17} Zhao, L. et al. 
	\textit{Nat. Phys.} \textbf{13}, 250-254 (2017).
	
	\bibitem{KAM13} Kampfrath, T. et al. 
	\textit{Nat. Photon.} \textbf{7}, 680-690 (2013)

	\bibitem{YAN19A} Yang, X. et al. Ultrafast nonthermal terahertz electrodynamics and possible quantum energy transfer in the Nb$_3$Sn superconductor. \textit{Phys. Rev. B} \textbf{99}, 094504 (2018)
		
	\bibitem{MAT14} Matsunaga, R. et al. 
	\textit{Science} \textbf{345}, 1145-1149 (2014)

    \bibitem{CEA16} Cea, T. et al. 
    \textit{Phys. Rev. B} \textbf{93}, 180507(R) (2016)
    
    \bibitem{Manske} H. Krull, N. Bittner, G. S. Uhrig, D. Manske, and A. P. Schnyder 
    \textit{Nature Comm.} \textbf{7}, 11921 (2016), 
    
    
    
   \bibitem{CEA19} Udina, M. et al. 
   \textit{Phys. Rev. B} \textbf{100}, 165131 (2019)
   
   \bibitem{Kemper} Kumar A, and Kemper A.F., 
   \textit{Phys. Rev. B} \textbf{100},  174515 (2019)
    
   \bibitem{chu19}  Hao Chu et al. 
   \textit{arXiv:1901.06675} (2019)
        
	
	\bibitem{GIO19} Giorgianni, F. et al. 
	\textit{Nat. Phys.} \textbf{15}, 341-346 (2019)
	
	
	\bibitem{Rajasekaran} Rajasekaran, S., Okamoto, J., Mathey, L., Fechner, M., Thampy, V., et al. 
	\textit{Science} \textbf{359}, 575-579 (2018). 
	
	\bibitem{YAN18} Yang, X. et al. 
	\textit{Nat. Mater.} \textbf{17}, 586-591 (2018)
	
	\bibitem{YAN19B} Yang, X. et al. 
	\textit{Nat. Photon.} \textbf{13}, 707 (2019)
	
	\bibitem{current} Nakamura, S.,  Iida, Y., Murotani, Y., Matsunaga, R., Terai, H., et al. 
	\newblock {\em Phys. Rev. Lett} \textbf{122}, 257001 (2019)
	
	\bibitem{peter} Cui, T. et al., 
	\textit{Phys. Rev. B} \textbf{100}, 054504 (2019)
		
\bibitem{Stephen1965}
M.~J. Stephen.
\newblock {\em Phys. Rev.} \textbf{139}, A197--A205 (1965)

\bibitem{Wu2017}
T.~Yu and M.~W. Wu.
\newblock {\em Phys. Rev. B} \textbf{96}, 155311 (2017)

\bibitem{Aoki2017}
Yuta Murotani, Naoto Tsuji, and Hideo Aoki.
\newblock {\em Phys. Rev. B} \textbf{95}, 104503 (2017)

\bibitem{Cea2018}
T.~Cea, P.~Barone, C.~Castellani, and L.~Benfatto.
\newblock {\em Phys. Rev. B}  \textbf{97}, 094516 (2018)

\bibitem{Haug} 
Haug, H. and Jauho, A.P., 
  \newblock  {\em Quantum Kinetics in Transport and Optics of Semiconductors},
\newblock   Springer Series in Solid-State Sciences,
 Springer Berlin Heidelberg (2007).
\bibitem{Kozlov2011}
Victor~V. Kozlov, Nikolay~N. Rosanov, Costantino De~Angelis, and Stefan
  Wabnitz.
\newblock {\em Phys. Rev. A} \textbf{84}, 023818 (2011)

\bibitem{Aoki2015}
Naoto Tsuji and Hideo Aoki.
\newblock {\em Phys. Rev. B} \textbf{92}, 064508 (2015)

\bibitem{Sadigh1998}
B.~Sadigh and V.~Ozoli\ifmmode \mbox{\c{n}}\else
  \c{n}\fi{}\ifmmode~\check{s}\else \v{s}\fi{}.
\newblock {\em Phys. Rev. B} \text{57}, 2793--2800 (1998)

\bibitem{PADUANI2009}
C.~Paduani.
\newblock {\em Solid State Communications} \textbf{149}, 1269 -- 1273 (2009)	

\bibitem{more1} L. Luo et al., 
\textit{Nat. Commun.} \textbf{10}, 607 (2019).

\bibitem{more2} T. Li et al., 
Nature {\bf 496}, 69 (2013) 	

\bibitem{more3} X.Yang et al., 
\textit{Phys. Rev. Lett.} \textbf{121}, 267001 (2018) 

\bibitem{more4} Patz, A. \textit{et al.} 
\textit{Nat. Commun.} \textbf{5}, 3229 (2014).

\bibitem{Patz2018} Patz, A. et al. \textit{Phys. Rev. B} \textbf{95}, 165122 (2017)

\end{thebibliography}

\begin{acknowledgments} 
This work was supported by National Science Foundation 1905981. 
The THz Instrument was supported in part by National Science Foundation EECS 1611454.
Work at the University of Wisconsin was supported by the Department of Energy Office of Basic Energy Sciences under award number DE-FG02-06ER46327 (structural and electrical characterizations) and Department of Energy Grant no.  DE-SC100387-020 (sample growth).
Theory work at the University of Alabama, Birmingham was supported by the US Department of Energy under contract \# DE-SC0019137 (M.M and I.E.P)
and was made possible in part by a grant for high performance computing resources and technical support from the Alabama Supercomputer Authority.

	\normalsize{$\dagger$ Equal contribution}
	\end{acknowledgments}


\end{document}